\documentclass[fleqn,10pt]{wlscirep}
\usepackage[utf8]{inputenc}
\usepackage[T1]{fontenc}
\usepackage{multicol}
\newenvironment{Figure}
  {\par\medskip\noindent\minipage{\linewidth}}
  {\endminipage\par\medskip}

\usepackage{subfigure}

\begin{document}

\title{Data-Driven Construction of Age-Structured Contact Networks}

\author[a,b,1]{Luke Murray Kearney \href{https://orcid.org/0009-0005-4553-1408}{Orcid}}
\author[b,c]{Emma Davis}
\author[b,d]{Matt Keeling} 

\affil[a]{MathSys Centre for Doctoral Training, Mathematics Institute, and School of Life Sciences, University of Warwick, Coventry, UK}
\affil[b]{Zeeman Institute for Systems Biology and Infectious Disease Epidemiology Research (SBIDER), University of Warwick, Coventry, UK}
\affil[c]{Statistics Department, University of Warwick, Coventry, UK}
\affil[d]{Mathematics Institute and School of Life Sciences, University of Warwick, Coventry, UK}


\begin{abstract}
Capturing the structure of a population and characterising contacts within the population are key to reliable projections of infectious disease. Two main elements of population structure -- contact heterogeneity and age -- have been repeatedly demonstrated to be key in infection dynamics, yet are rarely combined. Regarding individuals as nodes and contacts as edges within a network provides a powerful and intuitive method to fully realise this population structure. While there are a few key examples of contact networks being measured explicitly, in general we need to construct the appropriate networks from individual-level data. Here, using data from social contact surveys, we develop a generic and robust algorithm to generate an extrapolated network that preserves both age-structured mixing and heterogeneity in the number of contacts. We then use these networks to simulate the spread of infection through the population, constrained to have a given basic reproduction number ($R_0$) and hence a given early growth rate. Given the over-dominant role that highly connected nodes (`superspreaders') would otherwise play in early dynamics, we scale transmission by the average duration of contacts, providing a better match to surveillance data for numbers of secondary cases. This network-based model shows that, for COVID-like parameters, including both heterogeneity and age-structure reduces both peak height and epidemic size compared to models that ignore heterogeneity. Our robust methodology therefore allows for the inclusion of the full wealth of data commonly collected by surveys but frequently overlooked to be incorporated into more realistic transmission models of infectious diseases.  
\end{abstract}


\maketitle
\section*{Introduction}
\begin{multicols}{2}
Mathematical modelling of infectious diseases has become an integral process in shaping public health response measures to epidemics and pandemic preparedness~\cite{brooks2021modelling,funk2020short, biggerstaff2016results, viboud2018rapidd}. Historically, epidemiological models assumed that the population of interest is homogeneous, or `well-mixed'. Under this assumption, all infectious individuals have an equal rate of transmission to any susceptible individual in the population. While this is an over-simplification, it has provided a robust and surprisingly accurate method of predicting infection dynamics and guiding public health decisions  
\cite{keeling2008modeling,dashtbali2021compartmental, molla2023mathematical, zhan2019real, salem2016mathematical}. 
In reality, transmission is frequently linked to proximity and social contacts, as exemplified by the commonly-used risk thresholds for COVID-19 transmission (being within 2m for 15 minutes)~\cite{ferretti2024digital}. Based on pioneering work from the social sciences~\cite{klovdahl1994social}, there has been a growing interest in capturing patterns of human social contacts and the network that is implied~\cite{eubank2004modelling, eames2015six} to inform infectious disease models.

When the edges (or links) of a network represent routes for possible transmission, caused by sexual contacts, social interactions or proximity, then the network embeds much of the important epidemiological information. In particular, the heterogeneity in network contacts (referred to as network degree) is linked to the heterogeneities in secondary case distribution recorded for many infections~\cite{galvani2005dimensions, lloyd2005superspreading, adam2020clustering, de2013largest, shen2004superspreading}. 
Networks can also capture other structures such as assortative mixing which amplifies the role of superspreaders~\cite{garnett1993contact, newman2002assortative}, clustering which enhances local transmission but reduces wider dissemination~\cite{keeling1999effects, volz2011effects} and long-range contacts which interconnect entire populations promoting rapid spread of infections ~\cite{watts1998collective, may2006network,danon2011networks}.

In general, complete data on population-level contact networks is infeasible to collect, although several attempts have been made. The use of electronic devices (wearable RFID sensors~\cite{salathe2010high,kiti2016quantifying} or Bluetooth enabled smartphones~\cite{leith2020coronavirus}) provides an automated method of data capture, but only informs about connections {\it within} the participating population. Contact data gathered from contact tracing of infected individuals~\cite{kleinman2020digital,ferretti2024digital} can also generate a network, but often only describes the realised transmission routes and frequently misses unknown (random) contacts. Instead, much of the information we possess about social contacts follows the foundational NATSAL~\cite{johnson1992sexual, mitchell2013sexual} and POLYMOD~\cite{mossong2008social} surveys, with researchers focusing on contact data from individual respondents -- ignoring how these contacts link within the wider population. Such surveys have been refined over time~\cite{danon2012social, gimma2022changes} and now provide a key component of epidemiological models; they have been collated into open source platforms like \href{https://socialcontactdata.org/}{socialcontactdata.org}, providing a standardized syntax for multiple survey data sets. 

In the majority of epidemiological modelling studies, the individual-level heterogeneity of contacts is ignored in favour of more general average patterns. Most commonly, the reported contacts have been used to determine age-structured mixing matrices, which provide information about the average level of contact between any two age groups~\cite{mossong2008social}. While this has provided the foundation for many important epidemiological studies~\cite{davies2020age,bedford2015global}, it neglects the clearly observable heterogeneity in contacts. The importance of this heterogeneity has long been recognised for sexual contacts and sexually transmitted infections~\cite{may1987commentary, anderson1986preliminary}, and has been rediscovered for network-based transmission~\cite{eames2002modeling}.

\begin{Figure}
    \includegraphics[width=\linewidth]{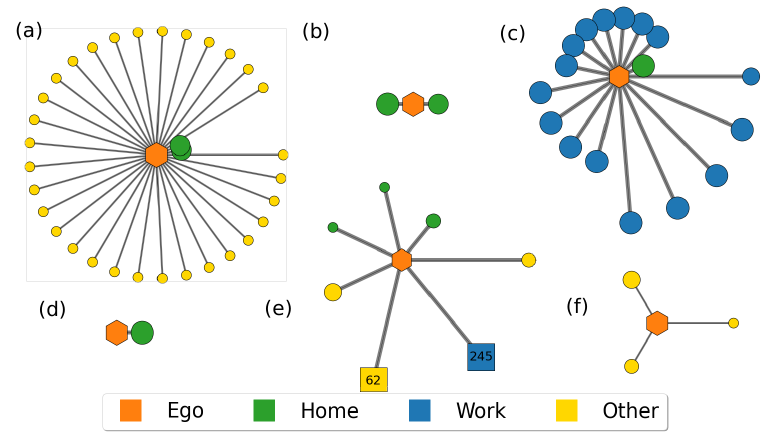}
    \captionof{figure}{Example participant ego-networks from CoMix~\cite{gimma2022changes}, showing individual heterogeneity. (a) School student, male 12-17, in lockdown easing period (schools open). (b) School student, male 12-17, during lockdown (schools closed). (c) Nurse, Male 20-29. (d) Mathematician, Female 30-39. (e) General Manager, Female 50-59. The `work' and `other' square nodes represent 245 and 62 short and infrequent contacts. (f) Retired, Female 70+. The participant (ego) is the orange central hexagonal node, connected circles represent individual contacts, and squares represent group contacts with a common location, duration and frequency. Node size represents contact duration. Edge length represents the frequency of social interaction with shorter lengths corresponding to longer contacts. Colours represent social settings of encounters (green, home; blue, work; yellow, other).}
    \label{fig:ego networks}
\end{Figure}

\begin{figure*}
    \centering

    \subfigure{}\hspace{\fill}\text{CoMix1}\hspace{\fill}
    
    \subfigure{}\includegraphics[width=0.36\linewidth]{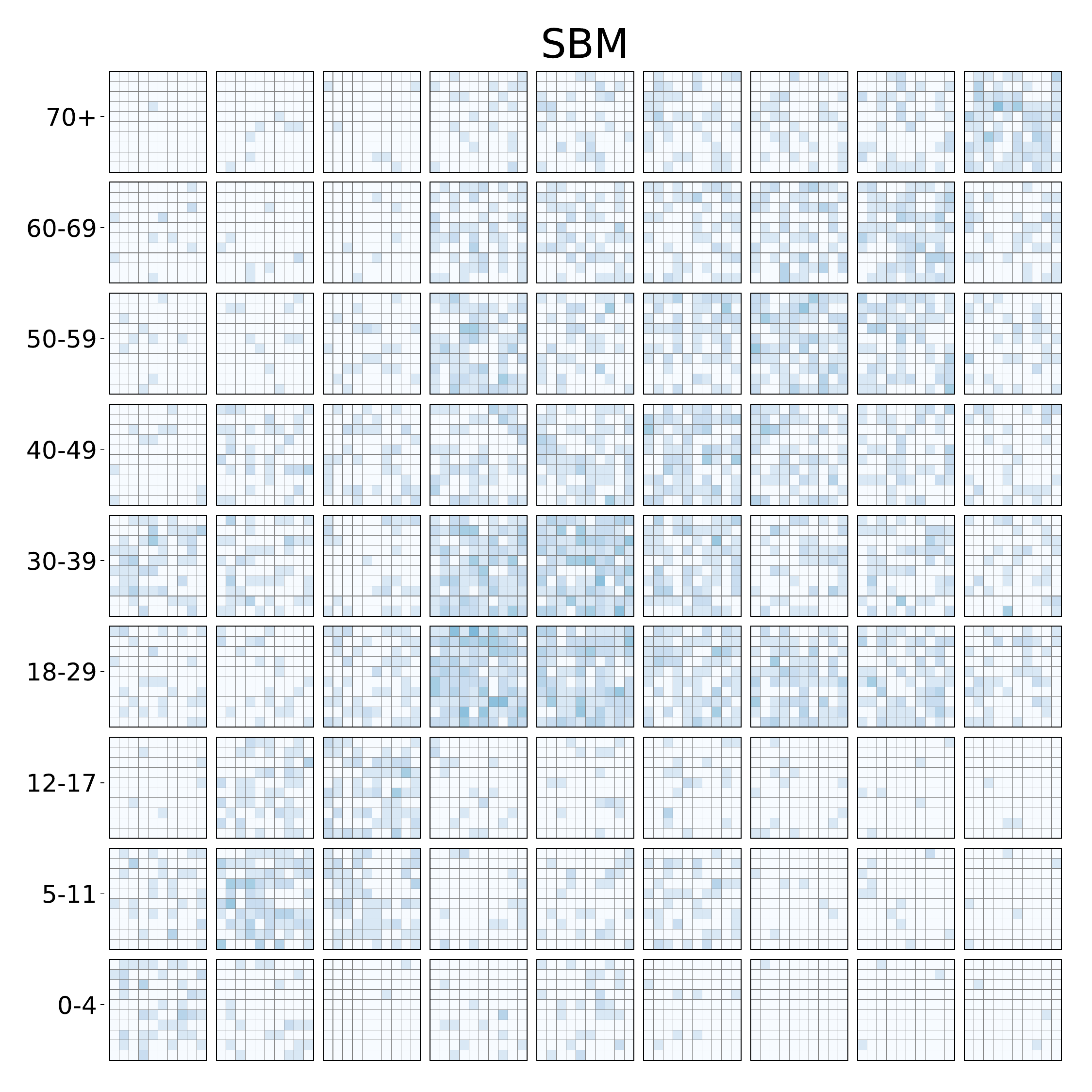}
    \subfigure{}\includegraphics[width=0.313\linewidth]{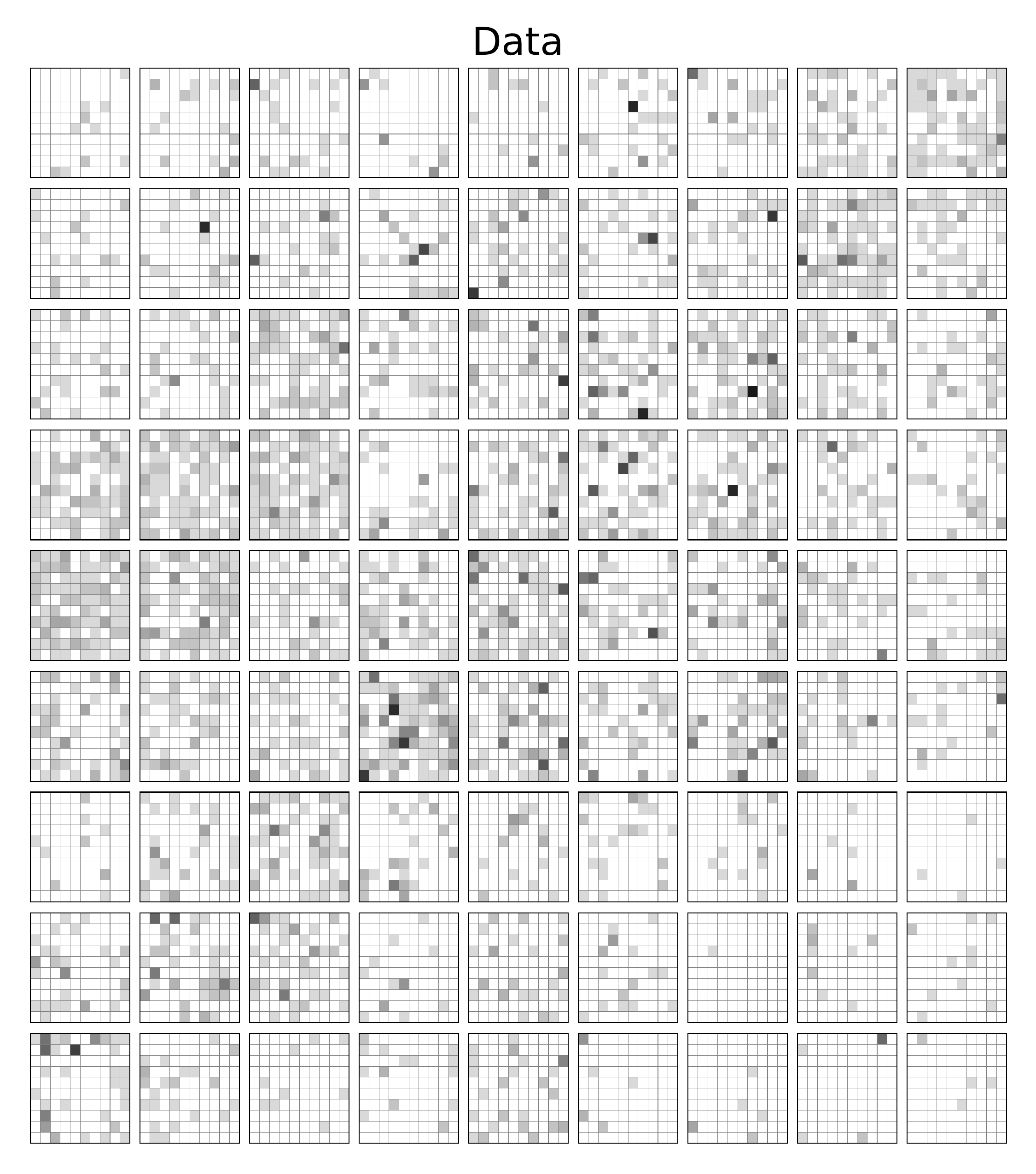}
    \subfigure{}\includegraphics[width=0.313\linewidth]{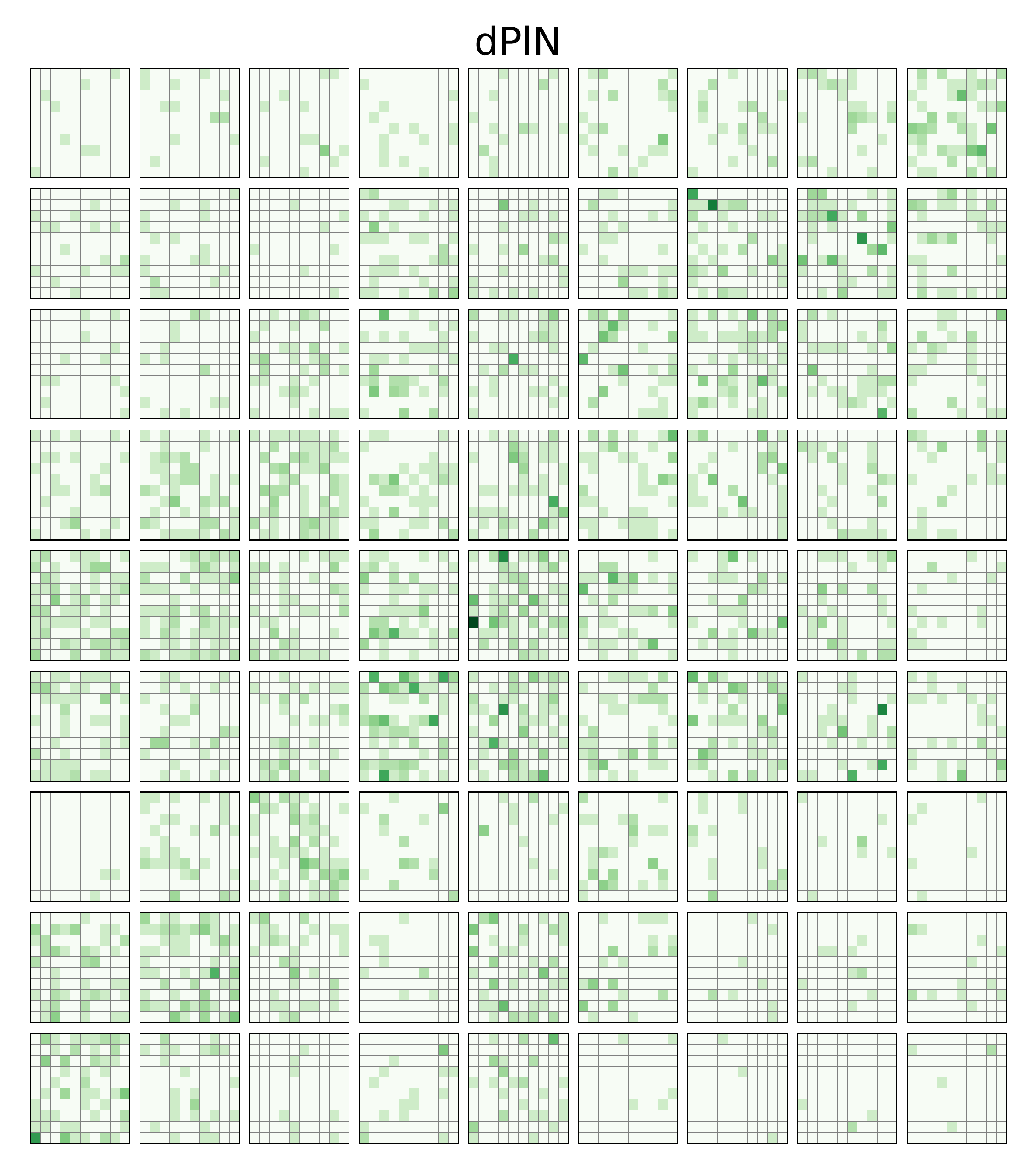}

    \subfigure{}\hspace{\fill}\text{CoMix2}\hspace{\fill}
    
    \subfigure{}\includegraphics[width=0.36\linewidth]{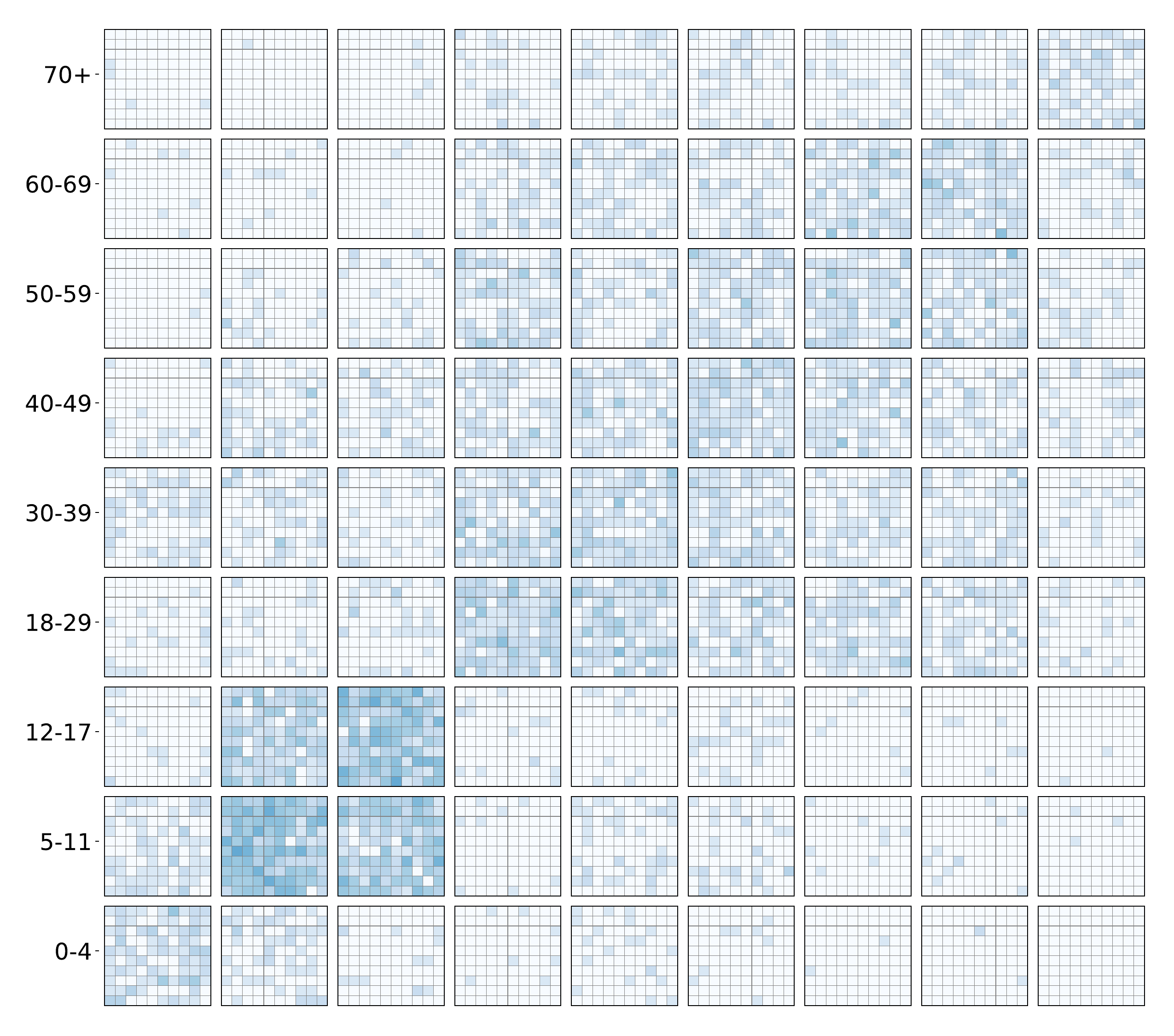}
    \subfigure{}\includegraphics[width=0.313\linewidth]{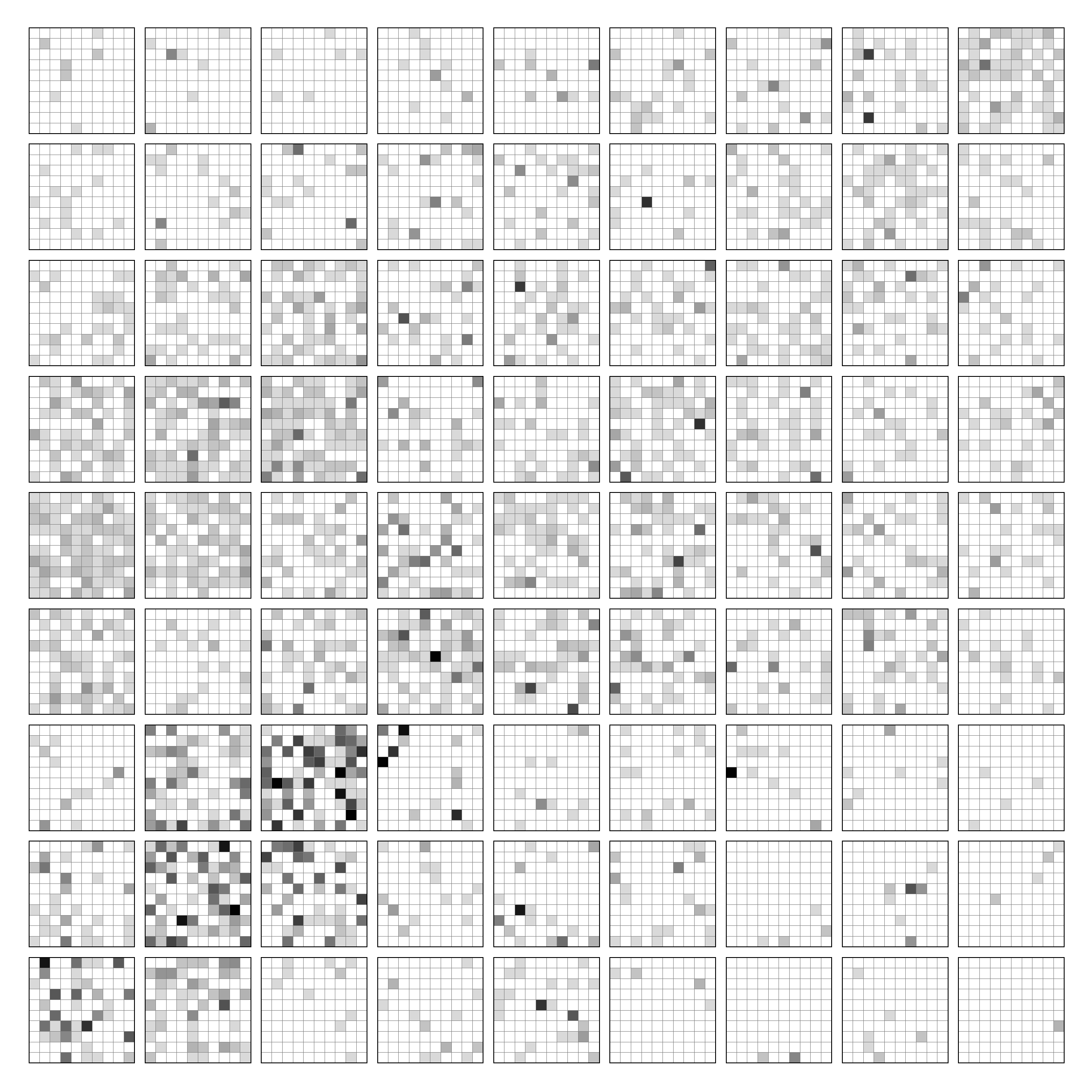}
    \subfigure{}\includegraphics[width=0.313\linewidth]{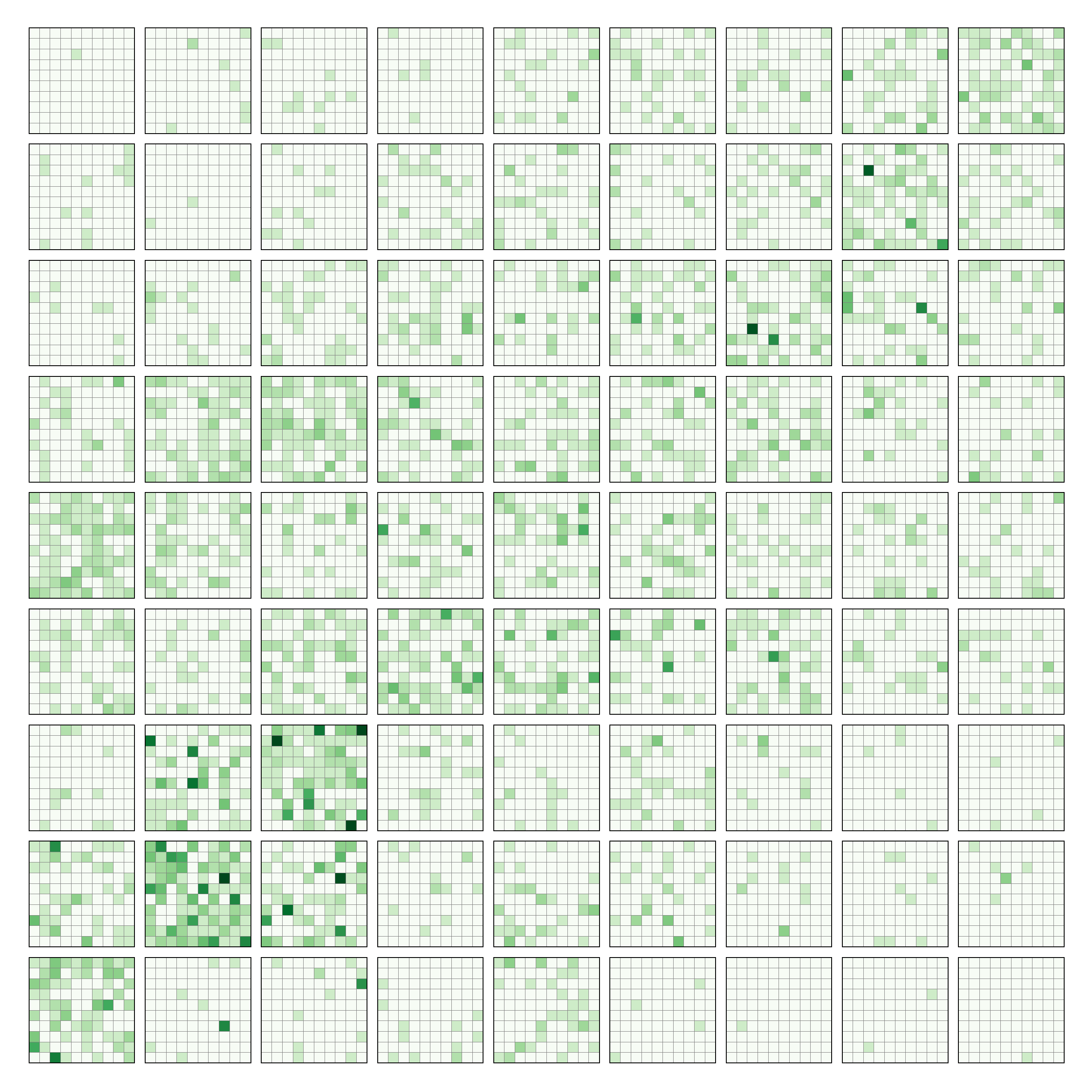}

    \subfigure{}\hspace{\fill}\text{POLYMOD}\hspace{\fill}
    
    \subfigure{}\includegraphics[width=0.36\linewidth]{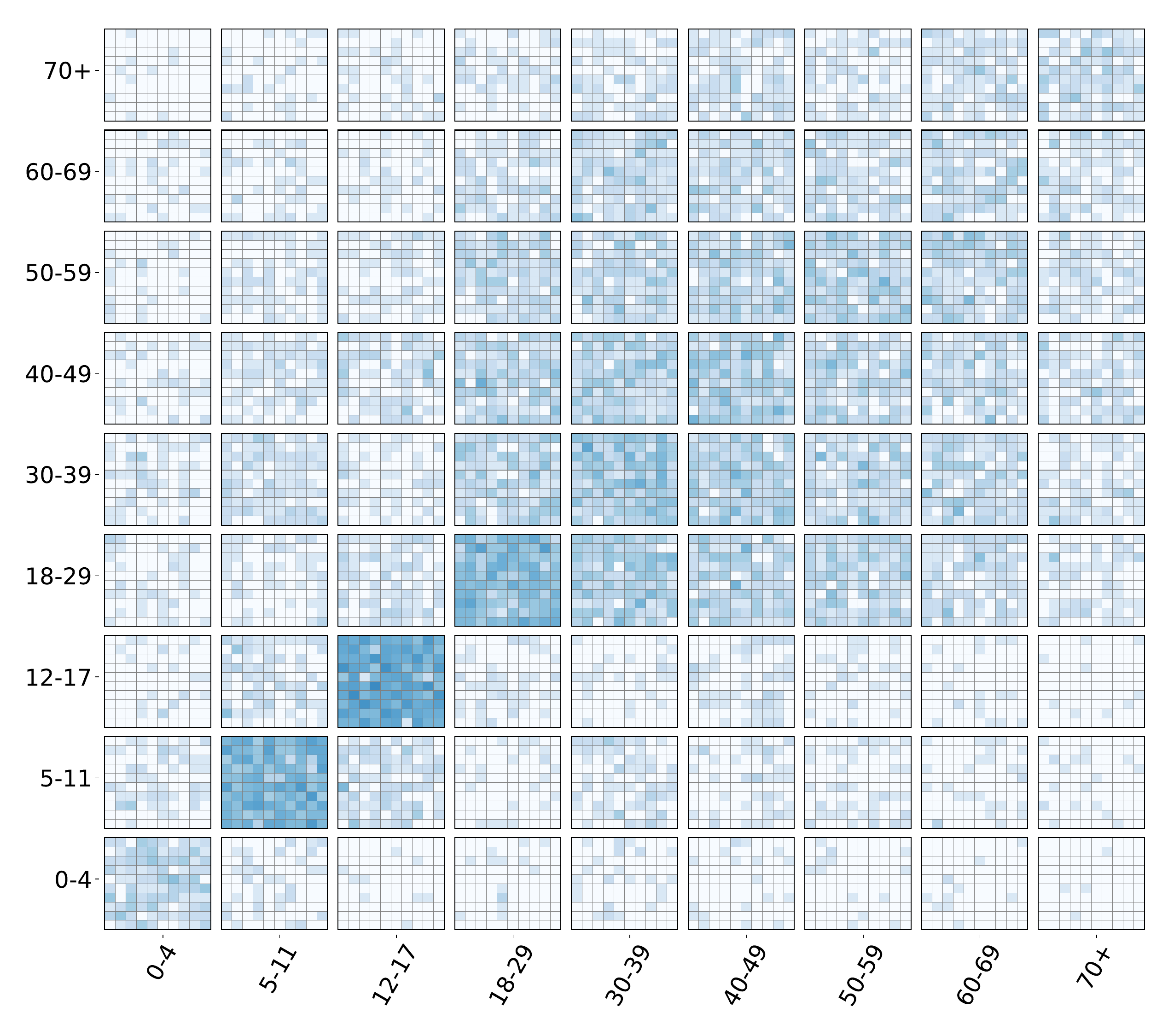}
    \subfigure{}\includegraphics[width=0.313\linewidth]{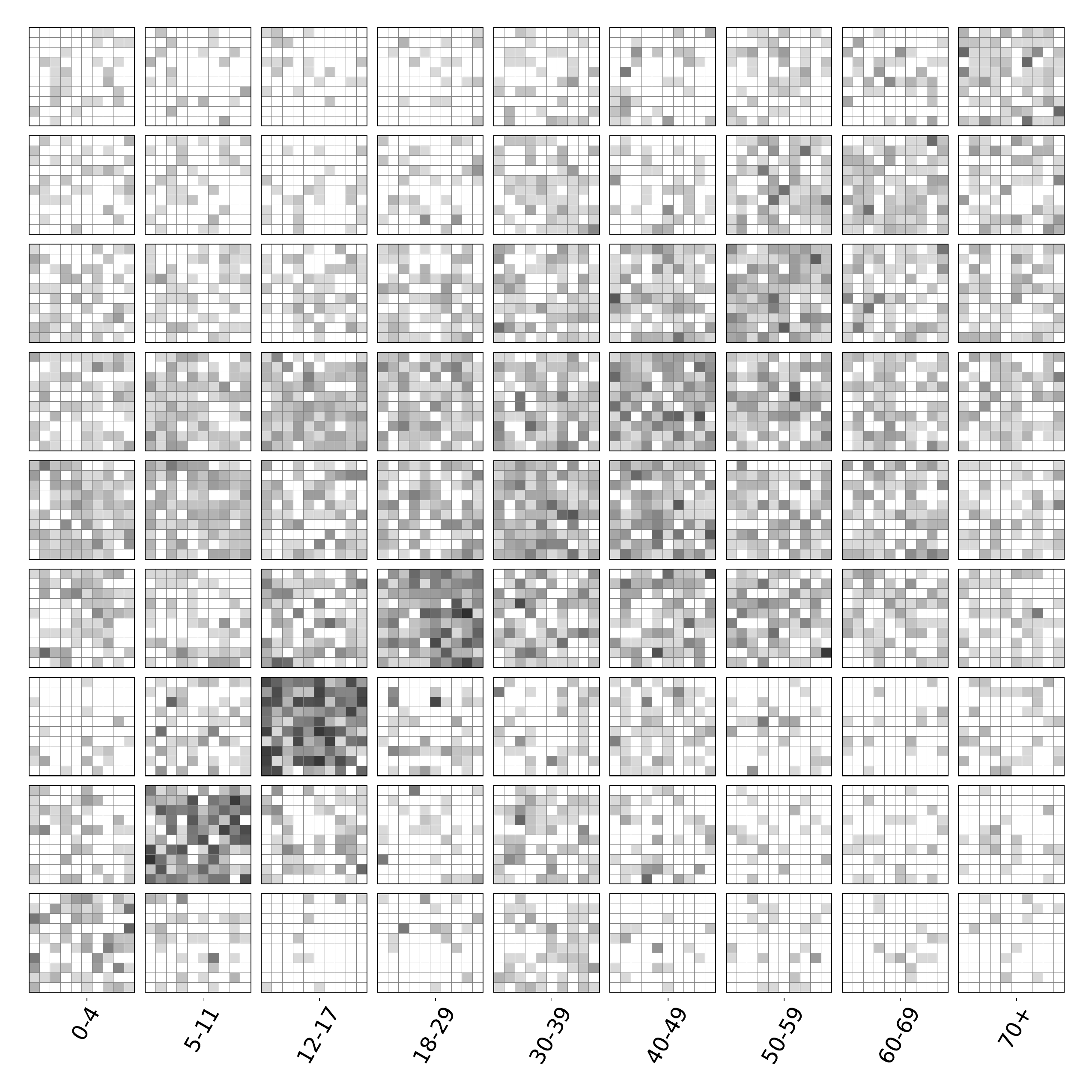}
    \subfigure{}\includegraphics[width=0.313\linewidth]{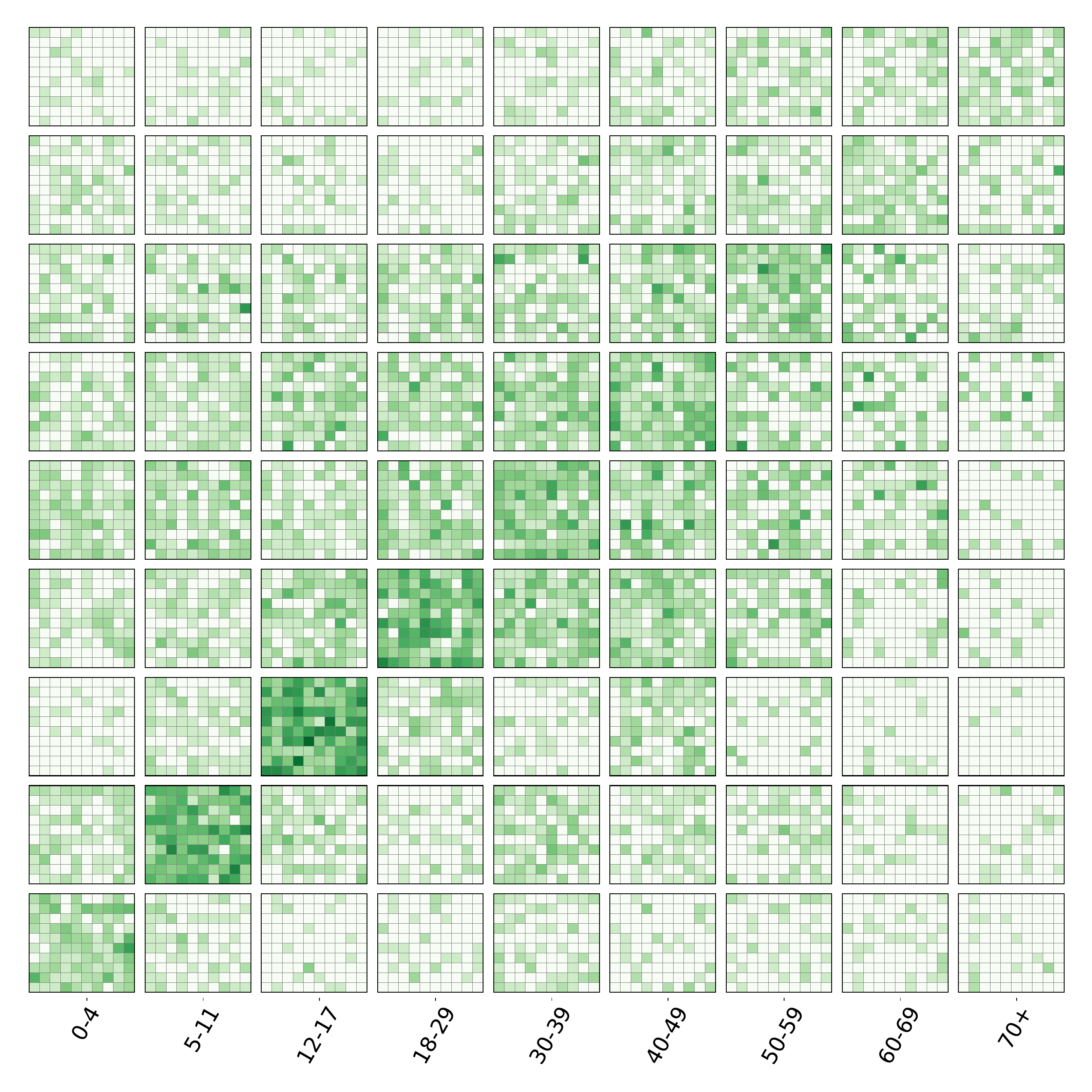}

    \hspace{\fill}\subfigure{}\includegraphics[width=0.26\linewidth]{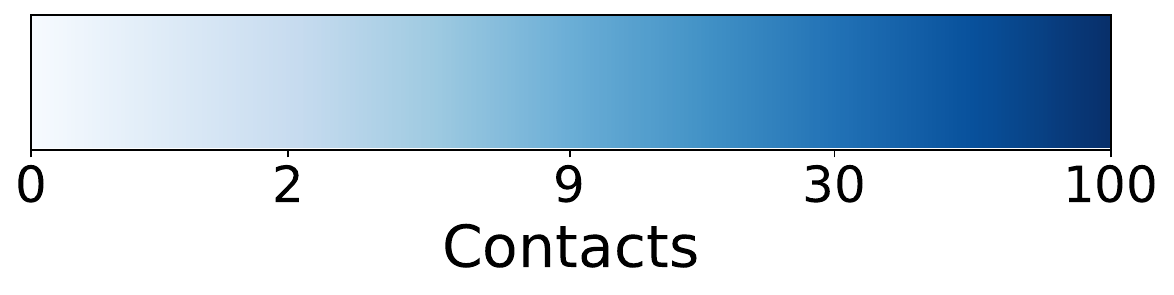}
    \hspace{\fill}\subfigure{}\includegraphics[width=0.26\linewidth]{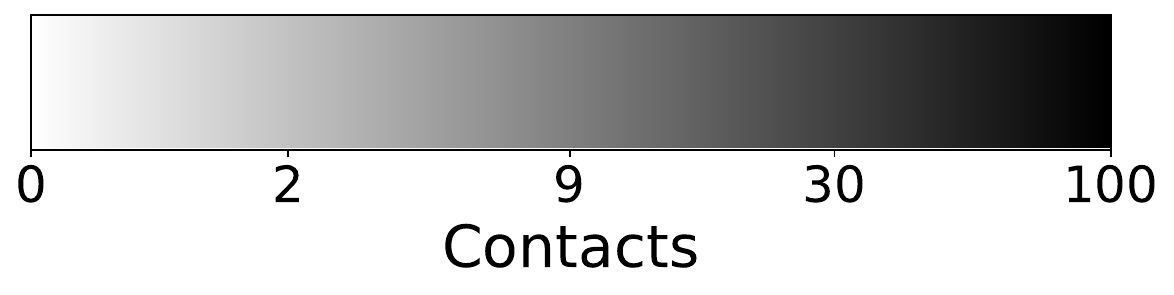}
    \hspace{\fill}\subfigure{}\includegraphics[width=0.26\linewidth]{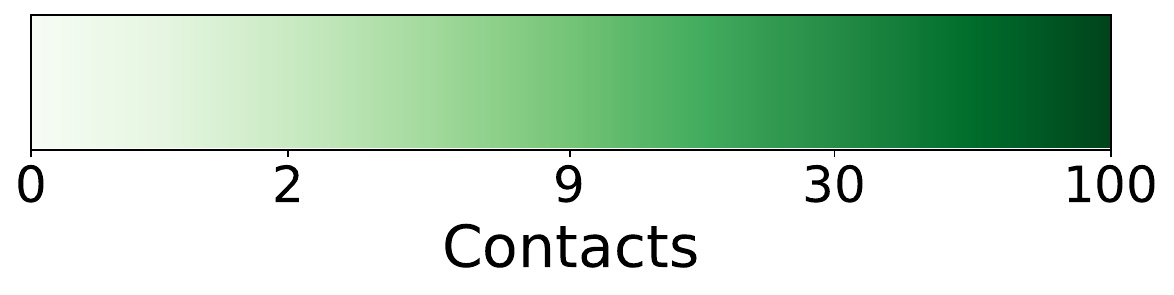}
    \hspace{\fill}
    
    \caption{Contact matrices representing the mixing between age-groups and highlighting the heterogeneities in the data (grey), the stochastic block model (blue) and the double Pareto log-Normal model (green). For the mixing between each pair of age-groups, we sample 100 ego-networks (associated with a respondant of the correct age) and calculate the number of contacts to individuals in the other age-group. The results are then plotted as a $10\times10$ subgrid to highlight the variability - points are colour-coded on a logarithmic scale (from 0 to 100) due to the extreme heterogeneities that are present.}
    \label{fig:contact matrices}
\end{figure*}

In this study, we formulate a novel method for the accurate reconstruction of age-structured networks from commonly collected survey data, that preserves both age-dependent mixing and contact heterogeneity. 
We demonstrate the power of this methodology on three data sets from the UK: the 2005/6 POLYMOD survey data~\cite{mossong2008social}; and CoMix data~\cite{gimma2022changes} from two different time periods in 2020 representing immediately post-lockdown ($30^{th}$ of July to the $3^{rd}$ of September 2020, referred to as CoMix1) and re-opening of schools ($4^{th}$ of September to the $26^{th}$ of October 2020, referred to as CoMix2).
Our method takes individual-level contact data from surveys together with a categorical classification of the respondent and contact (here taken to be 9 distinct age groups, but gender, occupation or sexual identity would be equally feasible), resamples the data (assuming negative binomial or power-law degree distributions for the number of contacts) to generate a larger synthetic population, and finally connects individuals to form a network. This network (and associated epidemic simulations) is then compared to the stochastic block model~\cite{holland1983stochastic} which preserves age-structure but not heterogeneity, a simpler version of our method which preserves heterogeneity but ignores age-structure, and classical homogeneous models which ignore both.

We compare our realised network to the underlying survey data using a generalisation of the Wasserstein distance measure, known as the  Earth Mover's Distance~\cite{rubner1998metric}, demonstrating that our method can construct networks that are closer to the data than existing methods. We then contrast epidemic simulations run on the three network formulations, using data from the three surveys, to consider the relationships between early growth, peak height of an epidemic and the final size of an outbreak. This highlights the profound impact of contact heterogeneity, even when the individual transmission rate is scaled to account for reductions in average contact duration with increasing number of contacts. We therefore conclude that existing age-structured models commonly ignore much of the information that survey data provide, potentially leading to erroneous epidemiological projections.


\section*{Results}

\subsection*{Network Model} 
We let $\mathbb{G}$ denote a network of $n$ nodes and associated connecting edges, defined by its creation method and the underlying data set (throughout we set $n=100,000$). Each node is classified into one of a finite number of groups -- in this instance an age-group from 1 through 9, representing people of age \{0-4, 5-11, 12-17, 18-29, 30-39, 40-49, 50-59, 60-69, 70+\} years. We extrapolate $\mathbb{G}$ from the survey data through a four-step methodology we refer to as the Heterogeneous Block Model (HBM). (1) We generate $n$ nodes that match the classification of the underlying population, in our example the age-distribution of the UK. (2) For each node, based on its classification (age) we pick the degree $k$ from a fitted distribution (negative binomial or double Pareto log-Normal), creating $k$ unconnected stubs for that node. (3) For each of the $k$ stubs we associate a classification (age) of who the stub should ideally connect to, again based on the survey data. (4) Bipartite and standard configuration models~\cite{molloy1995critical,guillaume2006bipartite} are used to connect stubs to their targets, capturing age-structured mixing. If any stubs remain unconnected, the linking process restarts, allowing stubs to connect to nodes most similar to their target classification. The stochastic block model (SBM)~\cite{holland1983stochastic} (with the ``communities'' in this approach defined by age classes) is used as a homogeneous comparator to our network model, capturing the between-age mixing without a heterogeneous degree distribution. (Figure \ref{fig:contact matrices} shows examples of the underlying age-dependent mixing matrices for the SBM, the raw data and our approach, derived from three different data sets.)

\subsection*{Network Accuracy}
It is important to understand how closely our network captures the contact data that is used in its formulation. 
A network reconstruction error score is calculated by finding the average error between each participant in the data set, $\boldsymbol{d_i}$ against a counterpart in the model $\boldsymbol{m}_j$. (Given the model is an extrapolation of the data, we cannot necessarily find the ``same'' individual in both.) The Earth Mover's Distance (EMD) metric~\cite{rubner1998metric} underpins this error score by utilizing optimal transport to quantify the difference between two ego-networks - by considering the number of errors in the ages, or the number of contacts that need to be added or subtracted for the two ego-networks to be equal (see Supplementary Material).

We begin by taking a sample of ego-networks, $\boldsymbol{m}$, of the same size and with the same age-distribution as the survey data. We create the matrix of errors $\boldsymbol{E}$, where $E_{i,j}=\text{EMD}(\boldsymbol{d_i}, \boldsymbol{m_j})$ if the pair come from the same age group, otherwise $E_{i,j}=\infty$. The problem of total network error is now reduced to an optimal bipartite matching problem between the data and sample, with cost matrix $\boldsymbol{E}$. The final (minimal) error corresponds to the best one-to-one match between individuals in the model sample and the data.

In Figure \ref{fig:EMD errors}, the error is calculated for 100 replicates of each network building model and data set, and amalgamated to the average error per individual. Increased degree heterogeneity (from the negative binomial and double Pareto log-Normal HBMs) reduces the reconstruction error in all cases, with the Stochastic Block Model performing significantly worse across all three data sets -- with an error up to three times larger than our model. This gap is widest for the CoMix data-sets, whose survey format allowing for over 100 contacts in a day lead to a power-law distribution. In contrast, the POLYMOD study shows less heterogeneity and is more readily captured by a negative binomial distribution. The black dashed line represents the average error between pairs of networks reconstructed using the dPlN model.
\begin{Figure}[ht]
    \centering
    \includegraphics[width=\linewidth]{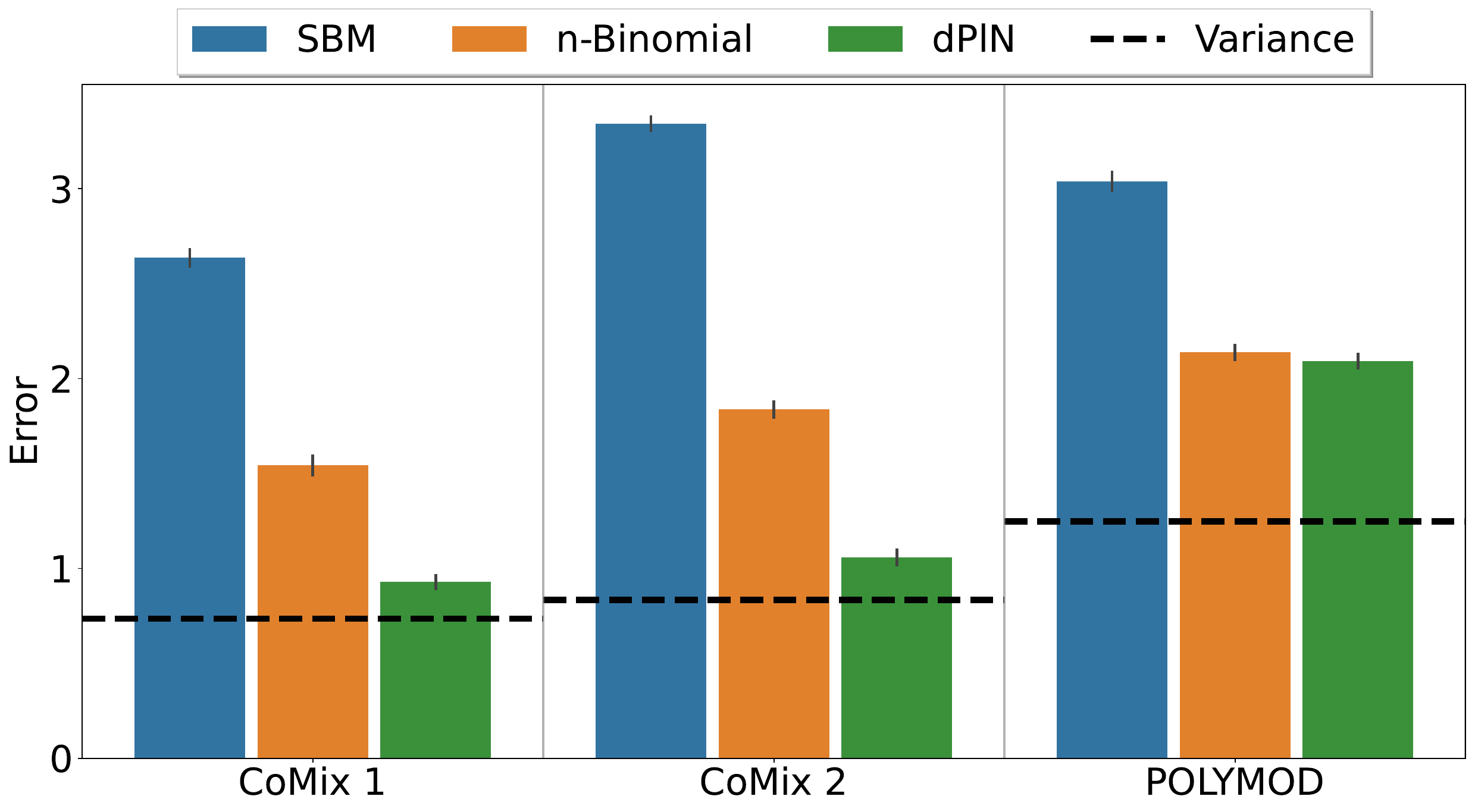}
    \captionof{figure}{Mean EMD error value per individual using the network construction methods for each data set, with (small) error bars of three standard deviations. Each network creation method creates a 100,000 node network 100 times, a representative sample of equal size to the data set is then compared to the data using EMD, giving an average error per person. The horizontal dashed line represents the variance in reconstructions of the same data, by calculating the the EMD between two networks constructed using our model.}
    \label{fig:EMD errors}
\end{Figure}
This quantifies the inherent error present when the network is built from a known dPlN distribution, akin to a minimum error when the underlying distribution is accurately captured. CoMix reconstructions are much closer to this minimum than POLYMOD (Figure \ref{fig:EMD errors}), again highlighting the differences in data caused by survey collection approach. 
The breakdown of errors in each age group Supplementary Material, Fig.S3 provides more information on where the difficulties in reconstruction lie and which age groups are better fit to these heterogeneous distributions. When schools are closed during lockdown in CoMix1 the disparity between the homogeneous and heterogeneous reconstructions are much smaller for age groups 5-11 and 12-17. This fact is also present in POLYMOD's 70+ age group where extra heterogeneity does not affect accuracy as strongly.

\subsection*{Simulation Model}
With information on which models best represent our network data, now we focus on how this increased realism affects epidemic simulation on the networks. Outbreaks of an SIR-type model are simulated using a Sellke construction~\cite{sellke1983asymptotic}, an exact methodology where all random numbers are sampled initially, including infectious durations ($T_i\sim \text{Exp}(\gamma^{-1}); \ \gamma=\frac{1}{5}$) and susceptiblity thresholds ($Q_i \sim \text{Exp}(1)$) for each individual. Given the time-varying force of infection:
\begin{equation*}
    \lambda_i(t)=\beta\sum_{j\in I(t)}A_{ij} f(\max(k_i,k_j)),
\end{equation*} 
individual $i$ becomes infected when the historical infection pressure ($\int_0^t \lambda_i(s) ds$)  surpasses the susceptibility threshold $Q_i$. Here $\beta$ is the infection rate parameter, $\boldsymbol{A}$ is the adjacency matrix of $G$ (which informs about connections between individuals $i$ and $j$) and $f(\max(k_i,k_j))$ is the transmission scaling of the link between $i$ and $j$. We have 2 different regimes for $f$: in the first we assume that all contacts are equally transmissible ($f\equiv1$); in the second we assume that transmission is based on the average duration of a contact ($f(k)=\overline{D}(k)$).  While we note that physicality, proximity and setting are all likely to influence transmission risk, we use the duration of contact as a parsimonious measure.  

The duration of a contact in CoMix is grouped into 5 discrete categories: less than 5 minutes, 5-14 minutes, 15-59 minutes, 1-4 hours and 4+ hours. 
To incorporate duration in constructed networks, the recorded average number of hours per contact is fitted to a function of the participant's degree (Figure \ref{fig:duration fitting}) using the functional form:
\begin{equation}
    \overline{D}(k) = Ae^{-Bk}k^{2} + Ck^{-E} + Fk^{-1},\label{eq: duration equation}
\end{equation}
This choice of functional form (with $E\in\left[ 0,1\right]$) ensures that the total infectiousness of an individual ($k \overline{D}(k)$) increases with their number of contacts, $k$.
\noindent In both regimes, index cases are chosen with probability proportional to the degree and potential duration scaling ($\mathbb{P}(x_i \in I(0)) \propto k_i f(k_i)$. This degree-dependent introduction ensures that infection is initially distributed among individuals with the most contacts.

\begin{Figure}[t]
    \centering
    \includegraphics[width=\linewidth]{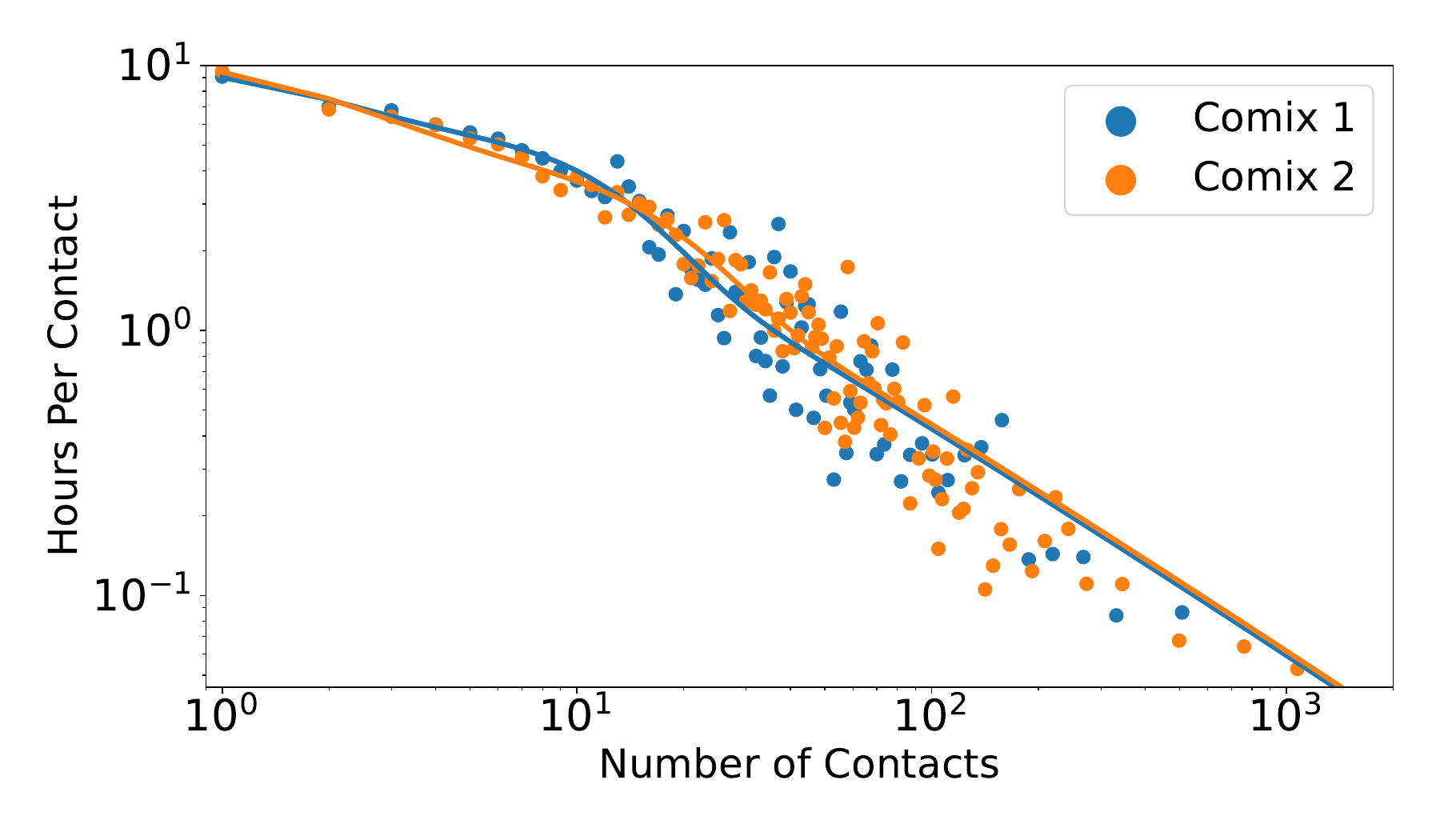}
    \captionof{figure}{For all respondents with a given number of contacts in the CoMix data sets, the average number of hours spent with each contact is plotted. Line of best fit added for the chosen functional form $\overline{D}(k)$ in Equation \ref{eq: duration equation}.} 
    \label{fig:duration fitting}
\end{Figure}

\begin{figure*}[!htb]
    \centering
    \subfigure{\includegraphics[width=.32\linewidth]{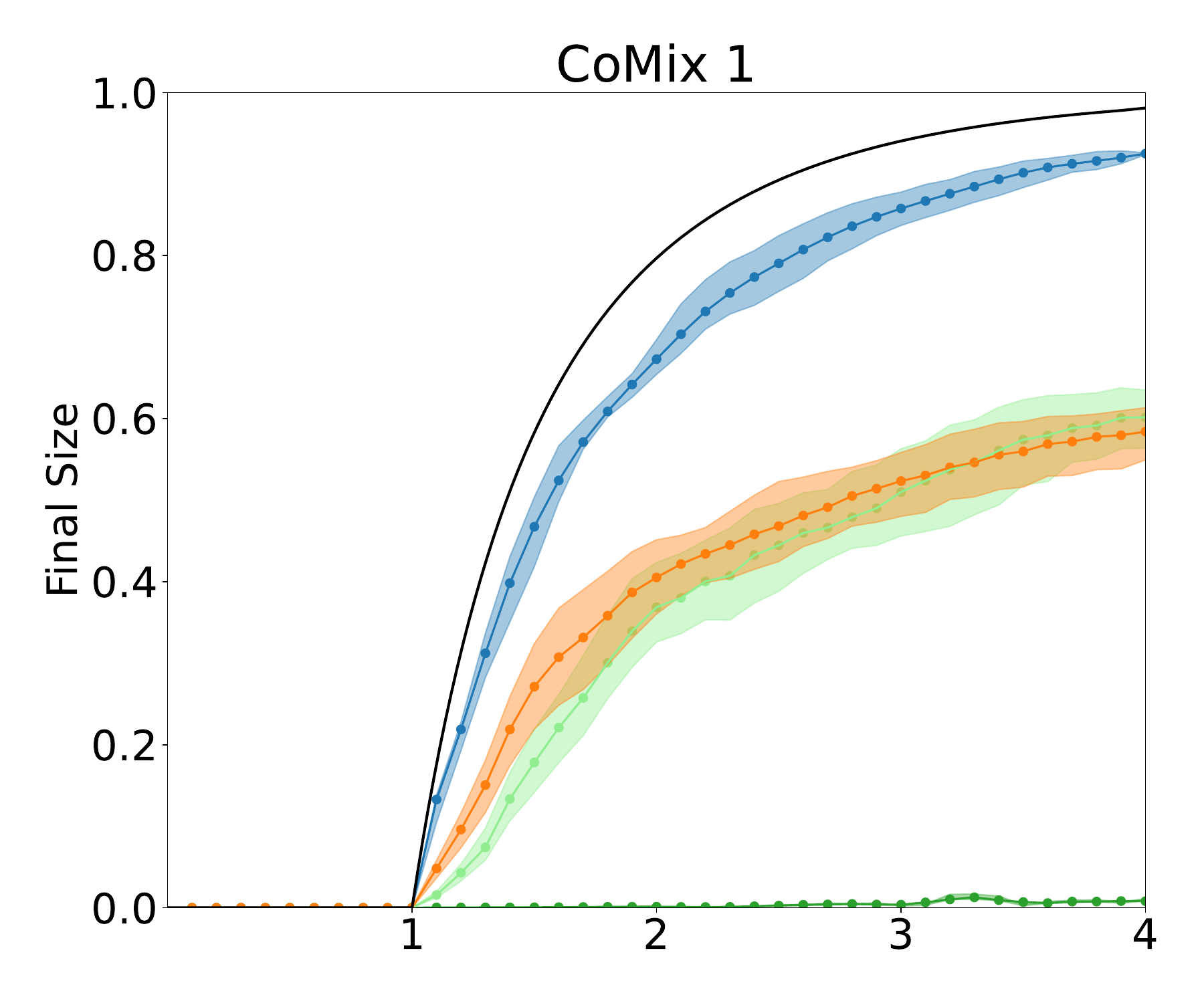}\label{fig:final size ribbon comix1}}
    \subfigure{\includegraphics[width=.32\linewidth]{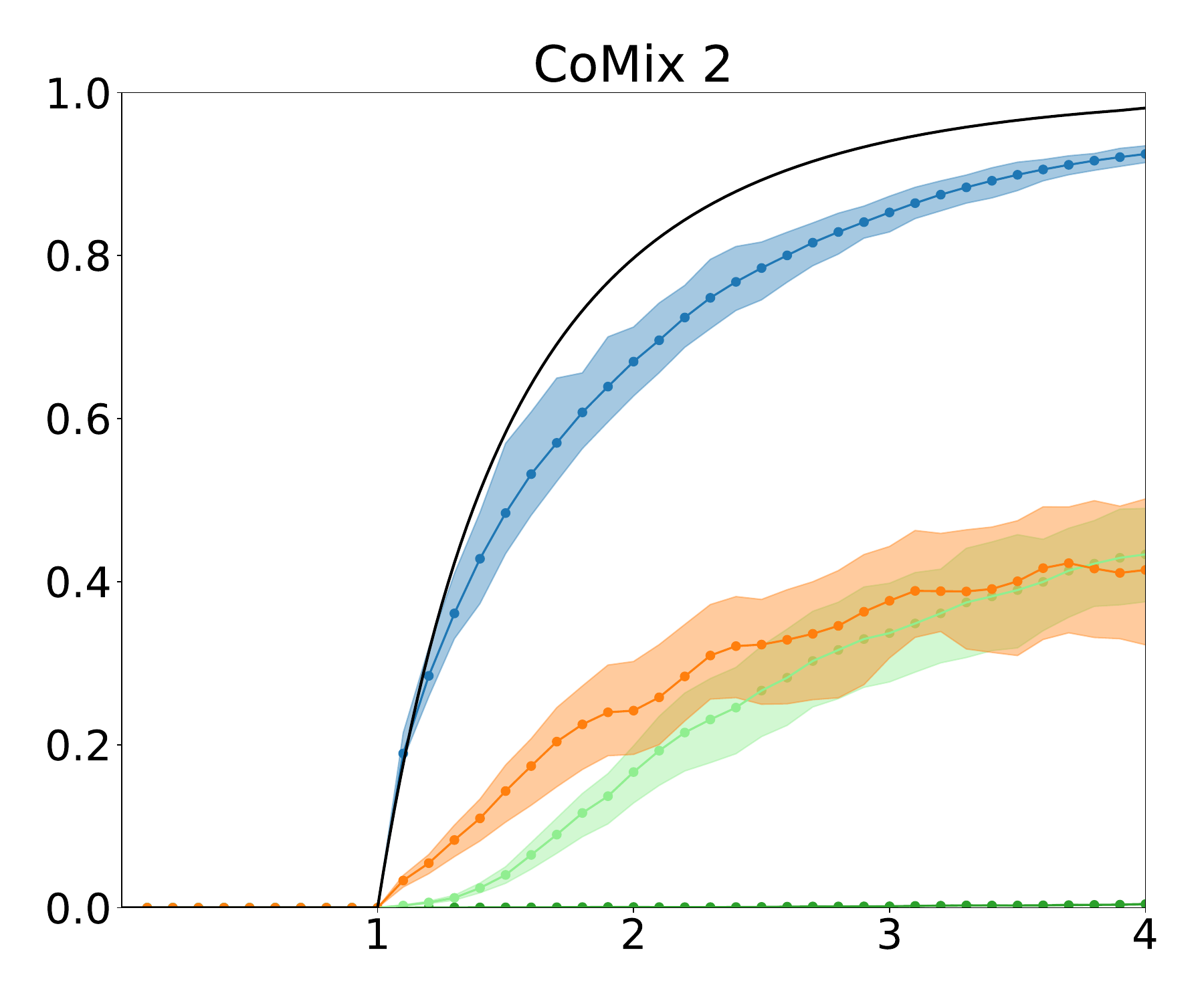}\label{fig:final size ribbon comix2}}
    \subfigure{\includegraphics[width=.32\linewidth]{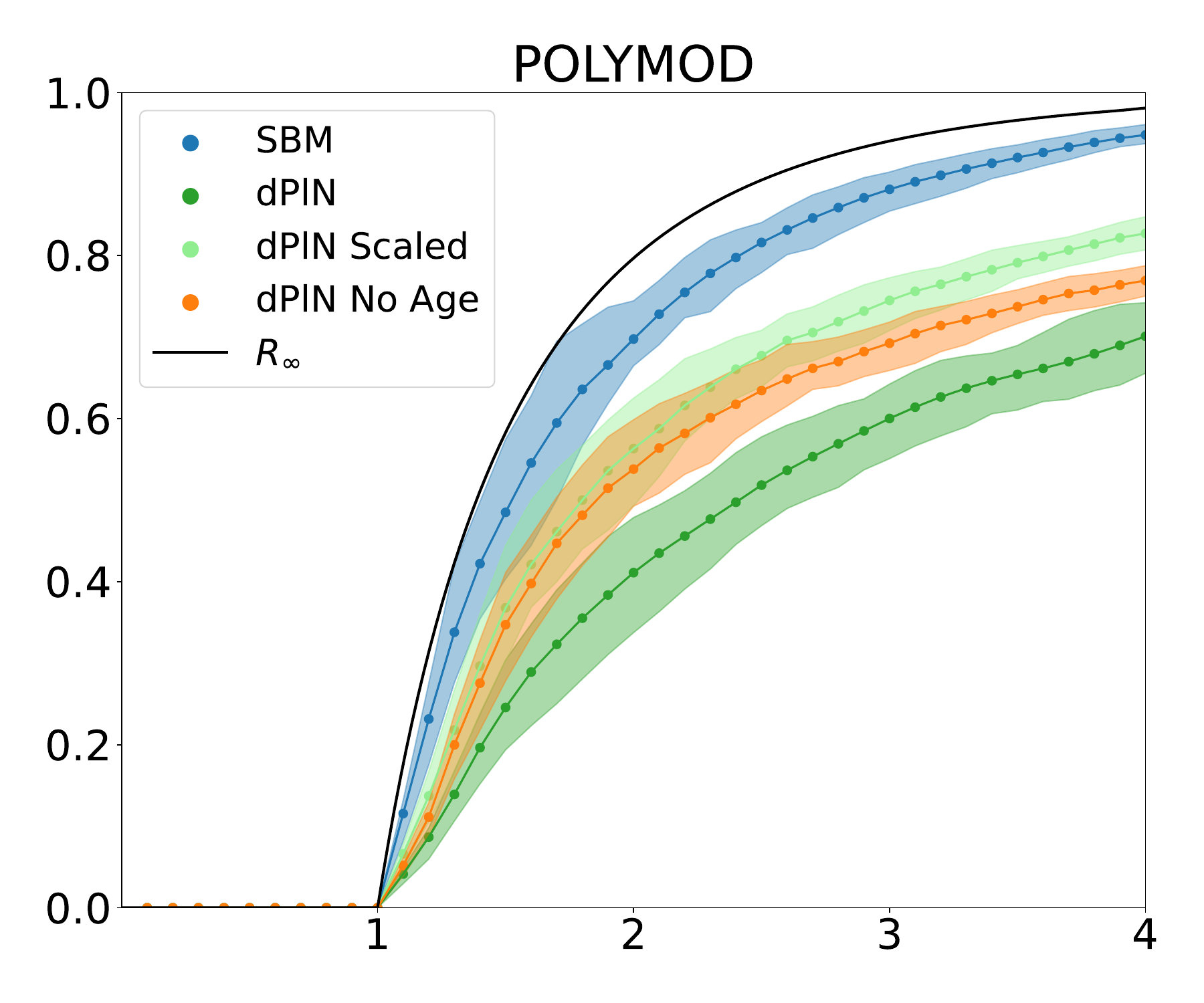}\label{fig:final size ribbon poly}}
    \subfigure{\includegraphics[width=.32\linewidth]{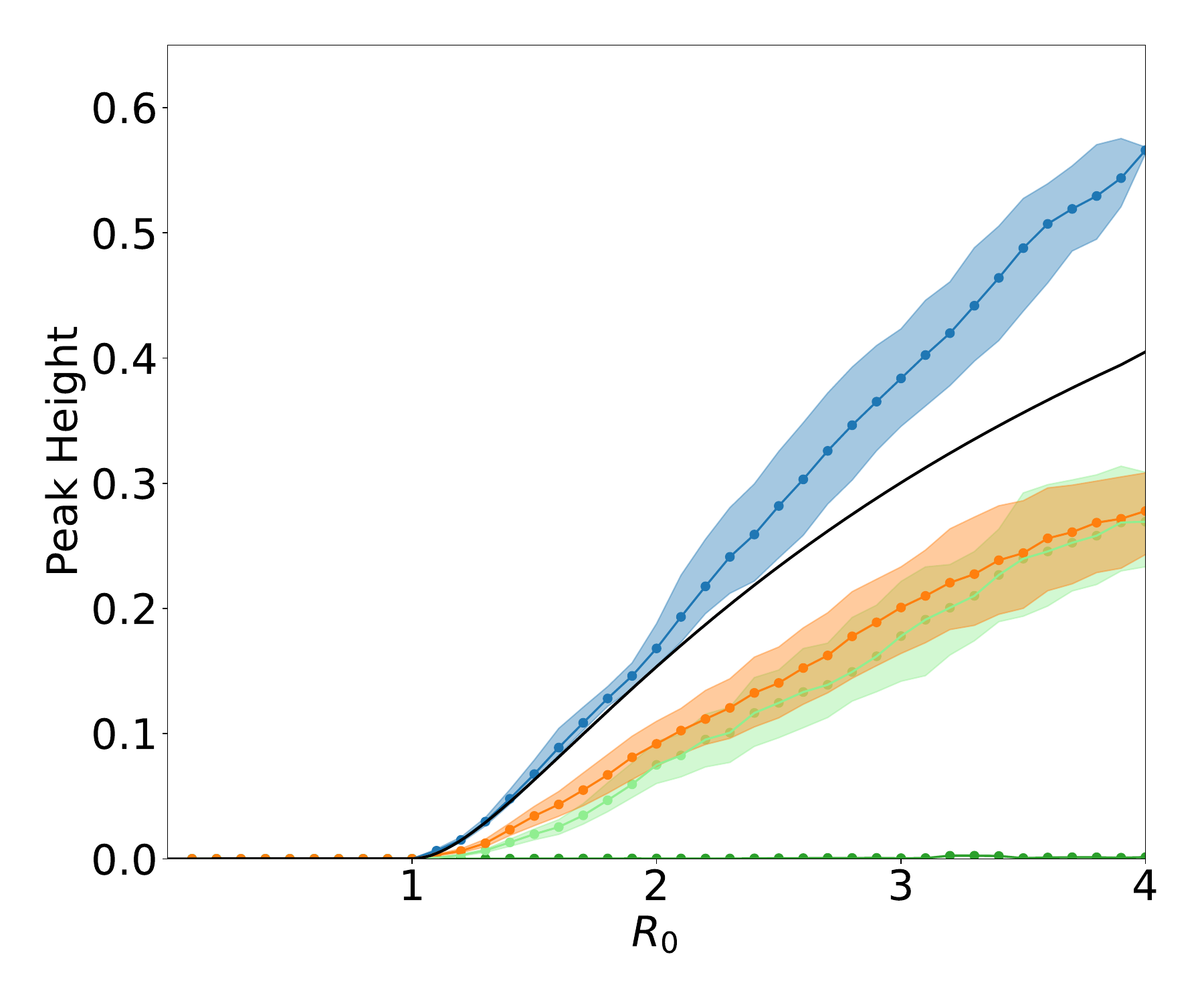}\label{fig:peak height ribbon comix1}}
    \subfigure{\includegraphics[width=.32\linewidth]{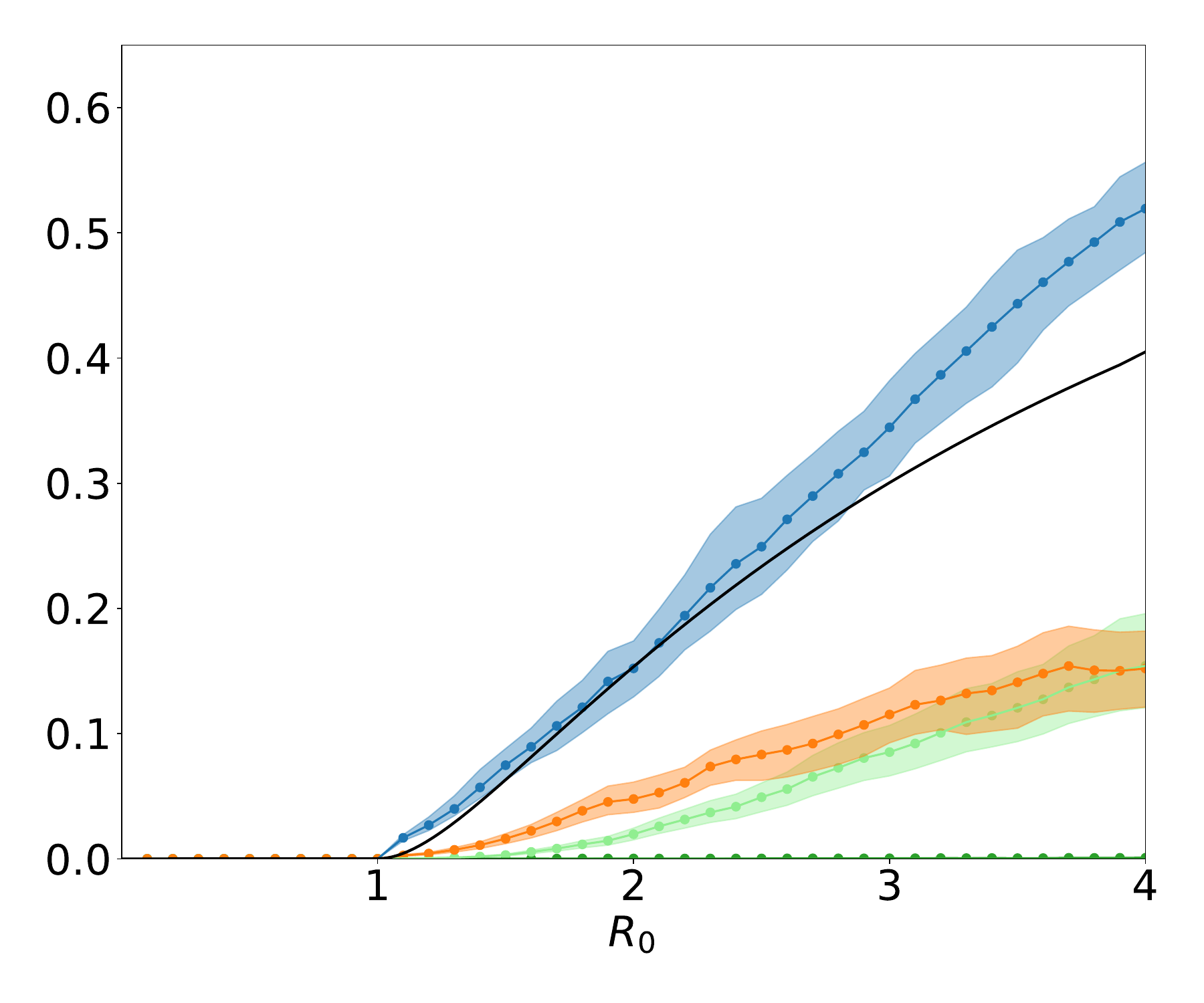}\label{fig:peak height ribbon comix2}}
    \subfigure{\includegraphics[width=.32\linewidth]{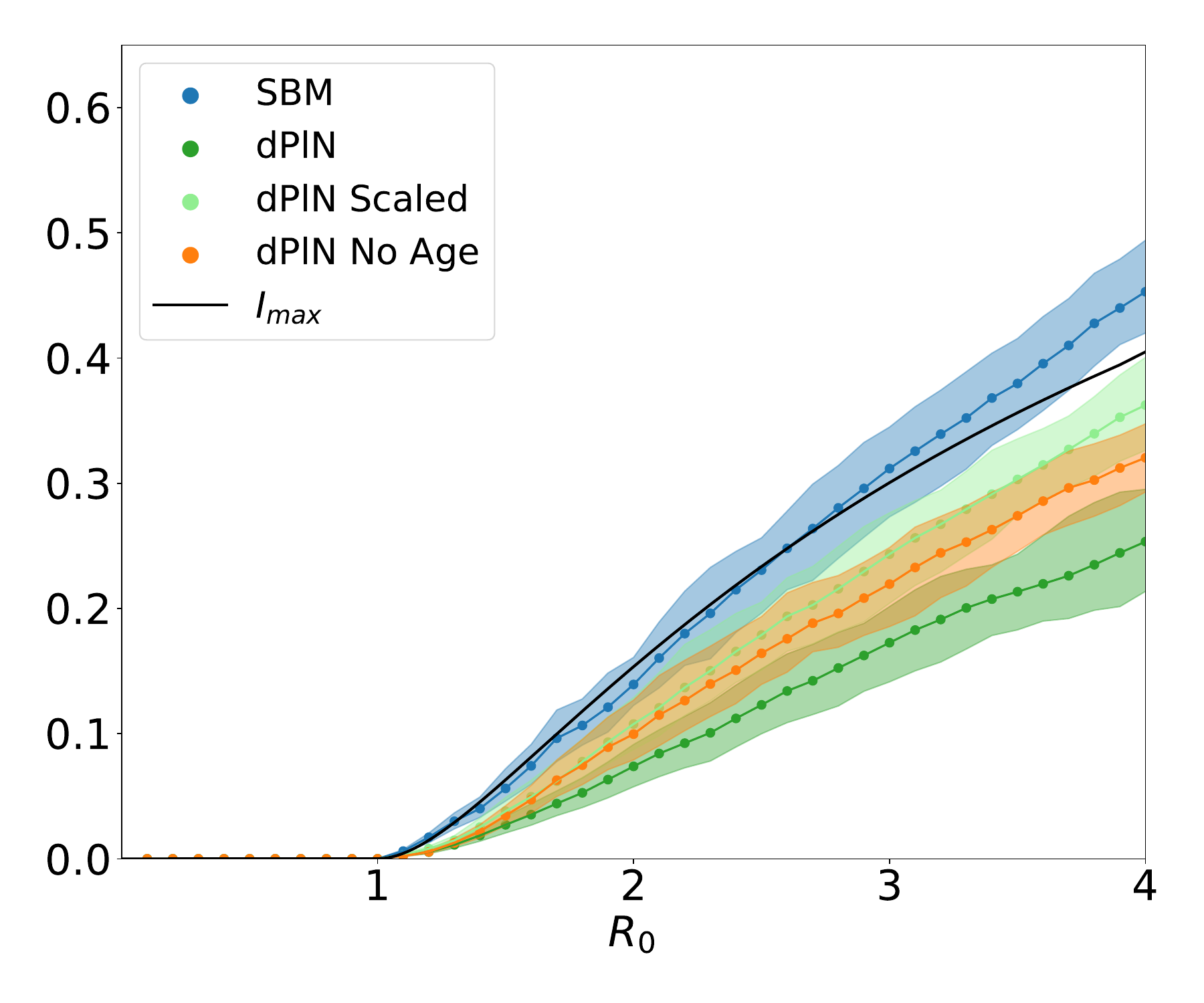}\label{fig:peak height ribbon poly}}
    
    \caption{(Row 1) The mean final size of each outbreak against the $R_0$ of that simulation, for SBM, dPlN unscaled, dPlN scaled and dPlN scaled without age-structured mixing. 95\% credible intervals are included for each model and the black line represents the theoretical final size of a deterministic ODE model. (Row 2) The peak height of the same simulations, with 95\% credible intervals. Here the black line represents the theoretical value.} 
    \label{fig:final size and peak height comparisons}
\end{figure*}

This simple simulation heuristic is designed to give an understanding of the relationship between network accuracy and prediction efficacy of our networks. In epidemiological modelling, the commonly used quantity $R_0$ refers to the average number of cases arising from a single infected individual, in a completely susceptible population. This can be used to characterise the early spread of an outbreak and to predict the scale of the outbreak from early outbreak data~\cite{van2002reproduction,de1994computation}. The precise relationship between $R_0$ and the final size of an outbreak, is known to be strongly dependent on the heterogeneity of transmission patterns~\cite{hebert2020beyond, elderd2013population}. Nevertheless, $R_0$ remains the most commonly used metric to characterize early outbreaks.

In our analysis the parameter $\beta$ is used to achieve the desired value of $R_0$, mirroring the standard fitting process adopted during the early stages of an outbreak. Precisely defining $R_0$ for a general network is a open problem, but as a proxy we use the average number of secondary cases infected by individuals in generation 1 (where generation 0 is the inital seeding of infection). 


In Figure \ref{fig:final size and peak height comparisons}, the final size and peak height of four network simulation-types with the same $R_0$ are compared: (i) a network generated by the stochastic block model (SBM - blue); (ii) and (iii) a network generated by the double Pareto log-Normal (dPlN - green) with
transmission constant across a connection (dark green) and with transmission scaled by the expected average duration (light green); and (iv) a double Pareto log-Normal network with duration scaled transmission that ignores age-structure (orange). (Note that given the lack of variability in the SBM, whether transmission is scaled or not has a negligable impact.)
The final outbreak size results for the SBM network are closest to but slightly below the theoretical final size ($R_{\infty}$) first proposed by Kermack and McKendrick~\cite{kermack1927contribution}. The other more heterogeneous networks, generated by the dPlN distributed HBM, lead to far smaller outbreaks. In particular, when all connections are assumed to be equally infectious (dPlN unscaled), the outbreaks are vanishingly small. This is because the early dynamics, which determine $R_0$, are set by rare highly-connected individuals; therefore while infection initially spreads rapidly it is unable to percolate through the bulk of the population. 
For the POLYMOD data, where very high numbers of contacts cannot be recorded, the four different simulations are far closer to each other, and even the unscaled dPlN network generates a significant outbreak. 

The peak height of the epidemic for the SBM is larger than the theoretical $I_{\max}$ due to the extra heterogeneity provided by the underlying network structure as compared to a completely homogeneous ODE model~\cite{barthelemy2005dynamical,volz2011effects}. Peak height for all other models lies below the theoretical prediction; the extra heterogeneity of these models which could increase peak height is outweighed by the substantially smaller final size of the epidemic, limiting how high the peak can be. 



Comparing CoMix1 when schools were closed, with CoMix 2 when schools had reopened, highlights our overarching message. When schools are open, the interaction between school-aged children dominates the mixing patterns (Figure \ref{fig:contact matrices}) and is the main contributor to determining $R_0$). 
Children mix intensely with each other, but weakly with the rest of the population. This disparity induces a strong early spread in schools, which quickly subsides in the sparsely connected exterior. Constraining $R_0$ in this regime reduces the final size by requiring a smaller $\beta$ to produce comparable early growth rates while $\sim 85\%$ of the population have not substantially increased their mixing. 


Finally, we consider the impact of keeping the heterogeneity of the dPlN network and retaining the duration of contact scaling, but removing the age-structure (orange points). These model projections are remarkably similar to those with full age-structure (light green points). We observe some differences for the CoMix2 data and low $R_0$ ($1<R_0<2$) and for POLYMOD and higher $R_0$ ($R_0>2$), which we attribute to the role of assortative mixing between children in these networks. 




Some quantitative assessment of the degree heterogeneity can be derived from the distribution of secondary cases per infected individual - observations of this distribution are often described by a negative binomial distribution~\cite{lloyd2005superspreading, endo2020estimating}. 
The negative binomial distribution is parameterised using the mean of the distribution ($R_0$) and the dispersion parameter, $\alpha$, which relates to the distribution variance: $\text{Var}(NB) = R_0 + R_0^2 / \alpha$. As such, low values of $\alpha$ are associated with high heterogeneity and importance of superspreading events. The dispersion parameter is highly dependent on both the disease and population. Estimates of $\alpha$ frequently lie in the range 0.1-0.7 for COVID-19~\cite{endo2020estimating, adam2020clustering, wang2021superspreading}, although it should be noted that these values come from fitting negative binomial distributions to relatively sparse and possibly incomplete data. 
Secondary case distributions taken from the early phase of our modelled outbreaks (with $R_0=2$) based on the CoMix 1 data are shown in Figure \ref{fig:secondary cases dist}. For each distribution we fit a negative binomial distribution using a least-squares method (dashed lines); comparing this to the reported range of $\alpha$ for COVID-19 provides a measure for each networks ability to recreate observed heterogeneity in transmission. 

The SBM network model (blue) generates a secondary case distribution with a shorter tail than observed (the best-fit negative binomial, $\alpha=0.87$), meaning that there are less super-spreading events than reported. Without scaling by the average duration, the dPlN network (dark green) overestimates the importance of highly connected individuals and hence under-estimates the dispersion parameter, $\alpha = 0.042$. However, even this unrealistically small value of $\alpha$ in the negative binomial distribution cannot capture the overdispersed nature of the model results.
The dampening effect of average duration scaling creates a secondary case distribution which lies between commonly reported dispersion parameters for COVID-19 (best fit $\alpha=0.47$).  These results again highlight the importance of individual-level data to support robust models that can fully capture the impact of observed heterogeneities.

\begin{Figure}
    \centering
    \includegraphics[width=\linewidth]{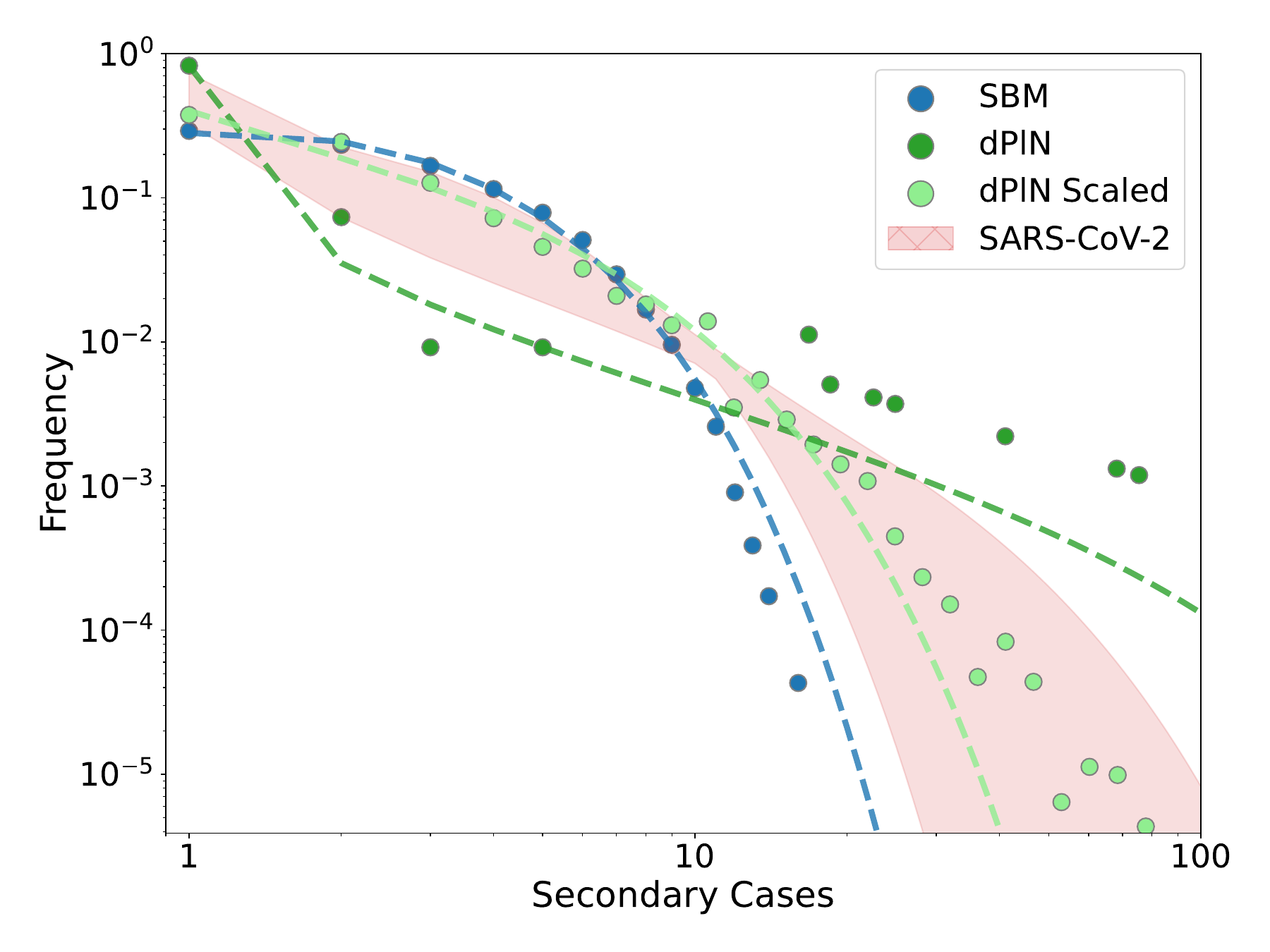}
    \captionof{figure}{Secondary case distributions with accompanying negative binomial fits for outbreaks simulated using the SBM model ($\alpha=0.87$), dPlN model ($\alpha=0.043$) and the dPlN scaled model ($\alpha=0.47$) for CoMix1. The shaded COVID-19 area represents a negative binomial distribution with $R_0=2$ and $\alpha\in[0.1,0.7]$ matching observations.}
    \label{fig:secondary cases dist}
\end{Figure}

\section*{Discussion}
Here we have created a general methodolgy (which we term the Heterogeneous Block Model: HBM) for the construction of age-structured contact networks from ego-centric network data (Figure \ref{fig:ego networks}). This methodology extrapolates from the available sample size and connects individuals to preserve the observed age-mixing patterns. Our extrapolation process allows us to capture the power-law distribution of contacts~\cite{adamic2001search,csanyi2004structure}. Throughout, we compare our network with that derived from the Stochastic Block Model (SBM)~\cite{holland1983stochastic} which ignores heterogeneity -- leading to Poisson degree distributions -- but retains age-structuring.


We have considered three exemplar data sets that provide ego-centric network information from the UK: the ground-breaking POLYMOD study~\cite{mossong2008social}; and two snapshots from the CoMix survey~\cite{gimma2022changes} taken during the COVID-19 pandemic. Our error metric -- comparing ego-networks from the data with those from our synthetic network -- highlights the need for a heterogeneous approach (compared to the SBM), and the power of using the double Pareto log-Normal (dPlN) distribution to capture the full distribution of contacts recorded in the CoMix data (Figure \ref{fig:EMD errors}). 
It should be noted that while these contact surveys are state-of-the-art in terms of quantifying human contacts, they are not necessarily a perfect reflection of epidemiologically important contacts. For example, they may suffer issues with recall-bias and estimating the age of contacts~\cite{mccormick2007adjusting,voelkle2012let}. Also, such surveys commonly record data on face-to-face conversational contacts, and while this is a good proxy for epidemic risk for infections spread through close contact, it misses long-term co-location which could play a role in long range air-borne transmission.


 
We simulated epidemic outbreaks on our networks based on their realised early growth, as captured by $R_0$,  and show that epidemic outcomes are heavily influenced by network heterogeneity. When all contacts are treated equally, the behaviour of the more accurate dPlN network is dominated by rare highly-connected superspreaders, who increase $R_0$ without leading to substantially larger outbreaks. In reality, the risk of transmission is likely to be positively correlated with the duration of a contact, and the CoMix data clearly shows that average duration declines with the number of contacts. By assuming that transmission risk is proportional to the average contact duration associated with a given degree, our model is able to capture the heterogeneous secondary case distribution that has been observed for COVID-19 and other infectious diseases and is often characterised using a negative binomial~\cite{danon2012social}. A wealth of other factors are likely to influence the risk of transmission across a contact including the intimacy of contact and the setting in which it occurs~\cite{ferretti2024digital}, but such information is difficult to gather and therefore hard to robustly include in modelling approaches.


The relationships between $R_0$ and epidemic size for the different models demonstrate 
how age-structure and degree heterogeneity shape an outbreak. We have shown that when considering the aggregate dynamics, age-structure plays a limited role (once simulations are matched to the same initial $R_0$). However, it is worth stressing that for many important infectious diseases (e.g. influenza or COVID-19) a population average is not a useful measure -- especially when disease severity is strongly age-dependent~\cite{davies2020age}, or when age-related interventions, such as school closures, require careful evaluation~\cite{andersen2020early}. For the example of COVID-19, being able to more accurately predict epidemic size and peak for specific age groups could support more accurate assessment of key public health outcomes, such as peak hospitalisations. Degree heterogeneity has a striking effect, even when it is moderated by the average duration -- which declines with increasing degree. The diversity between our results highlights how higher order network structures, not captured by $R_0$ (nor other early measures of epidemic growth) can profoundly impact the course of an outbreak. 

From a survey design perspective, our results demonstrate the need for robust survey designs that capture the full heterogeneous nature of social contacts~\cite{danon2011networks}. In particular, there is a clear difference in contact distributions reported by the CoMix~\cite{gimma2022changes} and POLYMOD~\cite{mossong2008social} surveys -- caused by an artificial cap of 100 daily contacts imposed in the POLYMOD survey. This is in comparison to several individuals in CoMix reporting incredibly high numbers (>1,000) contacts in a single day. Further research is required to understand what behaviours generate these self-reported highly-connected individuals and characterise the infectious disease transmission risk in these scenarios.

As with any modelling approach, we have made approximations to reality with associated limitations, including assuming a direct link between number of contacts and duration, and not accounting for how setting (e.g. home, work) may influence transmissibility. Our network construction method connects stubs with appropriate age-classes; with appropriate methods of building the sample population, and with greater computational expense, this approach could be extended to setting and duration. Such a network would inherently capture the duration associated with each connection (rather than applying averages) and would impart greater structure to the network. In addition, our formulation assumes that the ages of an individual's contacts are chosen at random (based on the age contact matrices), but often this structure is highly aggregated; for example, teachers mix with far more children than the average -- in addition, most of their work colleagues are also teachers who mix with more children, hinting at greater levels of structure.   
Our network building approach is unlikely to lead to clusters (triangle-forming contacts) within the network, yet we intuitively expect many clustered connections in household, work and leisure settings~\cite{danon2012social}. Including clustering in a data-driven way is extremely difficult, as it would require survey participants to estimate information about the contacts of their contacts \cite{danon2012social}. A complete picture of human social interaction would also need to include the dynamic nature of contacts, but data on changing contact patterns over time is extremely rare~\cite{beraud2015french}. Throughout, we have used COVID-like parameters as a motivating example, but have not included the rich epidemiology associated with this infection -- such as age-dependent severity and infectivity -- hence our results are representative of generic epidemiological dynamics rather than aiming to provide robust public health projections for COVID-19.

We have provided a standardised approach for creating, testing and simulating infectious disease outbreaks on accurate categorically-structured networks, as well as potentially informing the design of future contact surveys. Our analysis was limited to a set of UK data sets comprising differing periods of social restrictions, but there are a large and growing number of contact survey data sets available for use, many of which use the same syntax. The methodology described here could be applied to any of these data sets, allowing for out-of-the-box application to any population of choice.

\newpage


\end{multicols}
\end{document}